\documentclass{basi}
\usepackage{txfonts}
\usepackage{graphicx}
\usepackage[authoryear]{natbib}
\bibpunct{(}{)}{;}{a}{}{,}
\usepackage{longtable}
\makeatletter
\newcommand\tablesize{\@setfontsize\tablesize{8}{9}}
\makeatother
\begin{document}

\title[Further GMRT observations of the Lockman Hole at 610~MHz]{Further GMRT
       observations of the Lockman Hole at 610~MHz}
\author[T.~S.~Garn et~al.]{T.~S.~Garn,$^1$\thanks{deceased}
       D.~A.\ Green,$^2$\thanks{e-mail: {\tt D.A.Green@mrao.cam.ac.uk}}
       J.~M.~Riley$^2$ \& P.~Alexander\kern1pt$^{2,3}$\\
       $^1$SUPA, Institute for Astronomy, Royal Observatory of Edinburgh,
           Blackford Hill, Edinburgh EH9 3HJ,\\
       \ United Kingdom\\
       $^2$Astrophysics Group, Cavendish Laboratory,
           19 J.~J.~Thomson Avenue, Cambridge CB3 0HE,\\
       \ United Kingdom\\
       $^3$Kavli Institute for Cosmology Cambridge,
           Madingley Road, Cambridge CB3 0HA, United Kingdom}

\pubyear{2010}
\volume{00}
\pagerange{\pageref{firstpage}--\pageref{lastpage}}

\date{Received \today}

\maketitle

\label{firstpage}

\begin{abstract}
We present further observations of the Lockman Hole field, made with the Giant
Metrewave Radio Telescope at 610~MHz with a resolution of $6 \times
5$~arcsec$^2$. These complement our earlier observations of the central $\approx
5$~deg$^2$ by covering a further $\approx 8$~deg$^2$, with an r.m.s.\ noise down
to $\sim 80$~$\muup$Jy~beam$^{-1}$. A catalogue of 4934 radio sources is
presented.
\end{abstract}

\begin{keywords}
catalogues -- surveys -- radio continuum: galaxies
\end{keywords}

\section{Introduction}\label{s:intro}

The `Lockman Hole' is a region of very low {\sc Hi} column density
\citep{1986ApJ...302..432L} near $10^{\rm h}46^{\rm m}$, $58^\circ$ (J2000.0),
and is an ideal region for deep observations at X-ray wavelengths, due to low
absorption. As a consequence the field has also become a standard field for
deep observing campaigns in the optical, infrared and radio. Deep infrared
observations are available for an area of $\approx 14$~deg$^2$ of the Lockman
Hole region as part of the Spitzer Wide-area Infrared Extragalactic (SWIRE)
survey \citep{2003PASP..115..897L}. Parts of the Lockman Hole have also been
observed in several deep X-rays surveys, including: (i) the Chandra Lockman
Area North Survey (CLANS) and Chandra Large Area Synoptic X-Ray Survey
(CLASXS), which cover $\approx 0.8$ and $0.4$~deg$^2$ centred near $10^{\rm
h}46^{\rm m}$, $59^\circ 0'$ and $10^{\rm h}34^{\rm m}$, $57^\circ 30'$
respectively \citep{2004AJ....128.1501Y, 2008ApJS..179....1T,
2009ApJS..185..433W} -- and (ii) an XMM-Newton survey of $\approx 0.2$~deg$^2$
centred near $10^{\rm h}52^{\rm m}$, $57^\circ 20'$ \citep{2001A&A...365L..45H,
2008A&A...479..283B}. Various radio observations have been made of portions of
the Lockman Hole region, including very deep Very Large
Array (VLA) and Giant
Metrewave Radio Telescope (GMRT) radio surveys of
parts of the CLANS, CLASXS and XMM-Newton X-rays survey regions (e.g.\
\citealt{2003A&A...398..901C, 2005AJ....130.2019O, 2006MNRAS.371..963B,
2008MNRAS.385..893B, 2008AJ....136.1889O, 2009AJ....137.4846O,
2009MNRAS.397..281I}).

We have previously observed a central portion of the Lockman Hole field
(\citealt{2008MNRAS.387.1037G}, hereafter Paper~I) at 610~MHz with the GMRT
(see \citealt{2002IAUS..199..439R}). These observations covered $\approx
5$~deg$^2$ in the middle of the Lockman Hole, with twelve individual pointings,
which overlapped portions of the Chandra CLANS and XMM-Newton X-ray fields.
These observations had a resolution of $6 \times 5$~arcsec$^2$, at position
angle $+45^\circ$, with a typical r.m.s.\ noise of $\sim
60$~$\muup$Jy~beam$^{-1}$ in the centre of the pointings. A catalogue of 2845
sources detected in the field was presented. These observations were part of a
series of relatively deep, wide field observations of several SWIRE fields, and
of the Spitzer Extragalactic First Look Survey region (see
\citealt{2007MNRAS.376.1251G, 2008MNRAS.383...75G, 2009MNRAS.397.1101G}) .

Here we present further observations of the Lockman Hole field at 610~MHz with
the GMRT. These observations are not as deep as those available for some areas
of the Lockman Hole (see references above), but cover a further $\approx
8$~deg$^2$, completing coverage of the Chandra CLANS and XMM-Newton X-fields,
and also covering the CLASXS field. The observations and their data reduction
are described in Section~\ref{s:obs}, and the results -- including details of
the catalogue of 4934 sources -- are presented in Section~\ref{s:results}.

\begin{figure}
\centerline{\includegraphics[width=13.5cm,clip=]{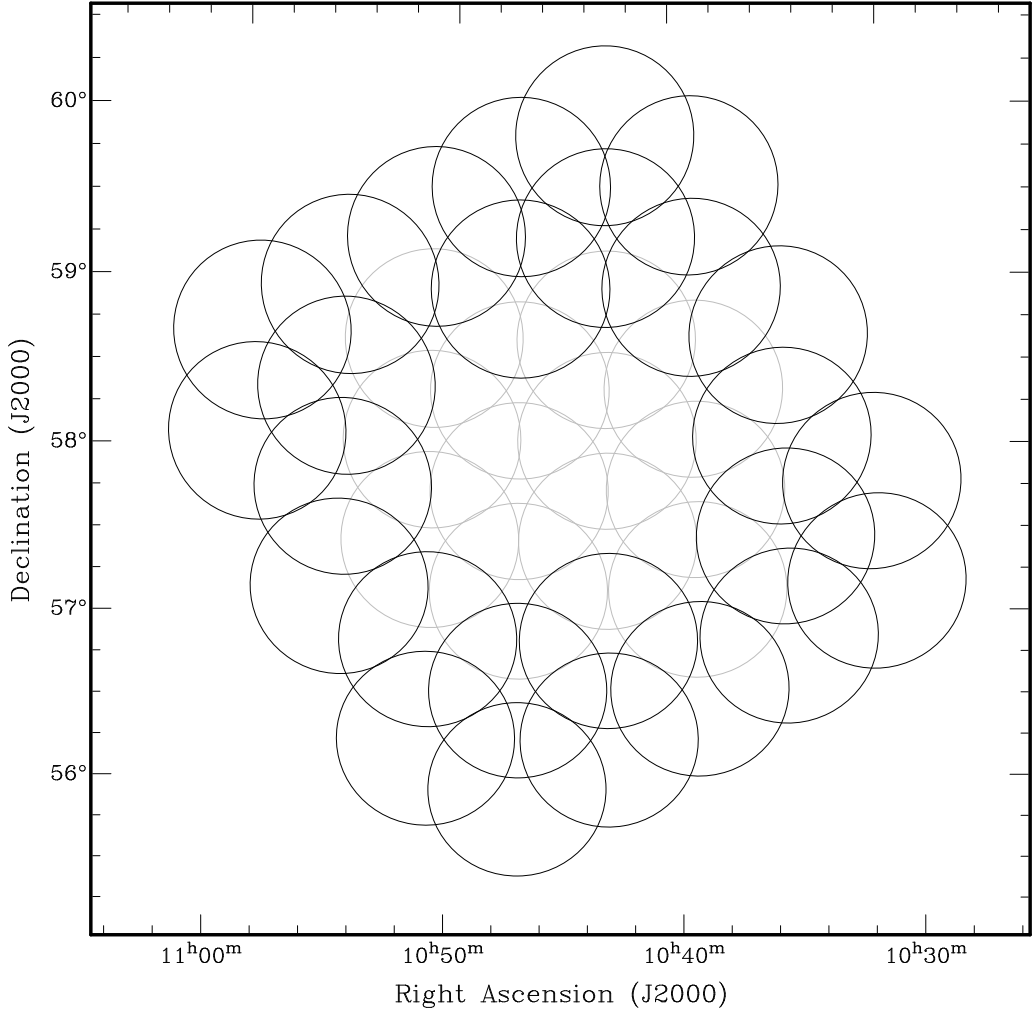}}
\caption{GMRT pointings in the Lockman Hole region, with circles indicating the
primary beam HPBW (the results from the central 12 pointings -- shown in grey
-- were presented in Paper~I).\label{fig:cover}}
\end{figure}

\section{Observations and Data Reduction}\label{s:obs}

The GMRT is an interferometer consisting of thirty 45-m antennas, twelve of
which are arranged within a central $\approx 1~{\rm km} \times 1~{\rm km}$
region, and the others in three arms giving baselines up to about 30~km. We
observed 26 pointings in the Lockman Hole region on a hexagonal grid
surrounding the 12 pointings observed in Paper~I (see Fig.~\ref{fig:cover}).
These pointings were initially observed on 2006 July 16 and 17, but the first
day's observations suffered from power outage in the eastern arm of the GMRT.
Consequently further observations were scheduled on September 7, 8 and 12, but
again power problems led to loss of some observing time, and a final
observation session was scheduled on October 4. In total, $\approx 21$~hours of
data, including calibration observations, were obtained. The observations were
made with two 16-MHz sidebands centred on 610~MHz, each of which was split in
128 narrow channels, with both left and right circular polarisations.

The observations consisted of interleaved scans of the different pointings,
typically $\approx 6$~min in duration, and observations of the nearby compact
calibrator source J1035$+$564 every 30 min or so. The flux density calibrators
3C48 or 3C286 were observed at the beginning and end of each observing session.
The data from each observing session were calibrated and processed using
similar procedures to those described in Paper~I. In summary: obvious
interference and other problematic data were flagged; the flux scale was tied
to 3C48 or 3C286, with assumed flux densities of 29.4 and 21.1~Jy respectively
at 610~MHz; the observations of 3C48 or 3C286 were used to characterise the
bandpass response of each antenna; the amplitude and phase stability of each
antenna were calibrated from the short observations of J1035$+$564. After the
calibrations were applied, the data were integrated into 11 channels of
bandwidth 1.25 MHz, and some further flagging of bad data was made. The
observations of each pointing, from all the different observation sessions, and
both sidebands, were combined; typically each pointing was observed for 5
scans, i.e.\ a total of about 30~min. For each pointing Stokes $I$ images were
then synthesised using 31 smaller facets, arranged in a hexagonal grid. All
images were synthesised with an elliptical restoring beam of size $6 \times
5$~arcsec$^2$, PA $+45^\circ$ (to match the images in Paper~I), with a pixel
size of 1.5~arcsec to ensure that the beam was well oversampled. The images
went through three iterations of phase self-calibration at 10, 3 and 1~min
intervals, and then a final iteration of phase and amplitude self-calibration,
at 10~min intervals, with the overall amplitude gain held constant in order not
to alter the flux density of sources. The self-calibration steps improved the
noise level by about 10~per~cent, and significantly reduced the residual
sidelobes around the brighter sources. The final r.m.s.\ noise, before
correction for the GMRT primary beam, varied considerably between pointings. In
particular, in the west of the field there is the bright source 3C244.1 (see
below), and the noise near this is increased, due to dynamic range limitations.
(Indeed, in Paper~I, the noise of the two northwestern inner pointings was also
noted as being high, due to the proximity of these pointings to 3C244.1.) The
noise in most of the pointings to the east is typically
80~$\muup$Jy~beam$^{-1}$, before primary beam correction. This is slightly
higher than for the inner 12 pointings presented in Paper~I, as expected due to
the somewhat shorter integration time per pointing. The r.m.s.\ noise values in
pointings in the west, particularly those close to 3C244.1 are considerably
worse, by factors of up to about three. Both the inner and outer pointings were
mosaicked together, with weighting appropriate to the relative noise of each
pointing. The contribution to the mosaic was cut off at the point where the
primary beam correction for each pointing dropped to 20 per cent, i.e.\ a
radius of 32~arcmin from the centre of each pointings.

\begin{figure}
\centerline{\includegraphics[width=13.5cm,clip=]{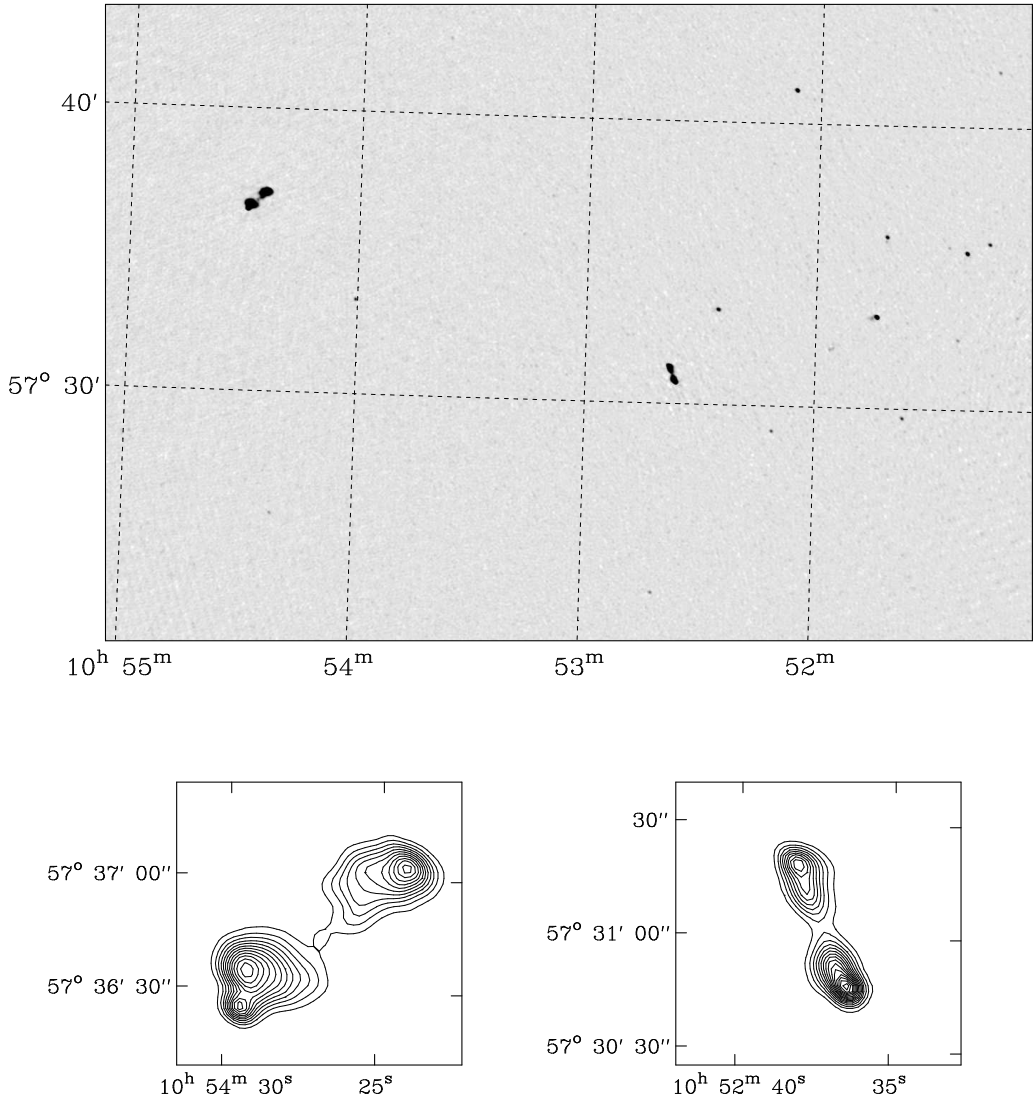}}
\caption{A sample 610-MHz greyscale image (from $-0.5$ to 4~mJy~beam$^{-1}$),
for an eastern outer portion of the Lockman Hole, with contour images for two
brighter double sources in the field (contour levels at 2, 4, 6$\dots$
mJy~beam$^{-1}$). The resolution is $6 \times 5$~arcsec$^2$, at a PA
$+45^\circ$.\label{fig:east}}
\end{figure}

\begin{figure}
\begin{tabular}{p{7cm}p{5.5cm}}
\raisebox{-5cm}{\includegraphics[width=7cm,clip=]{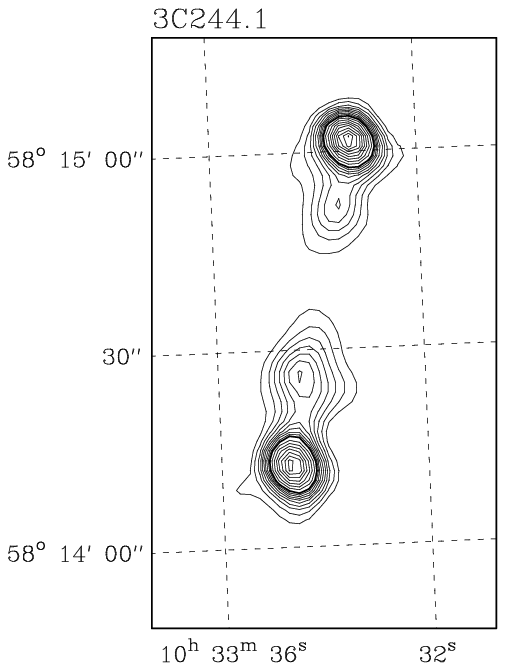}} &
\caption{610-MHz image of 3C244.1 from these observations. The contours are
0.1, 0.2, 0.3$\dots$ 1.0, 1.2, 1.4$\dots$ 2.8~Jy~beam$^{-1}$. The resolution is
$6 \times 5$~arcsec$^2$, at a PA $+45^\circ$.\label{fig:3C244p1}}
\end{tabular}
\end{figure}

\section{Results}\label{s:results}

The final mosaicked image is $13000 \times 13000$~pixel$^2$, so is difficult to
display in its entirety. A sample region in the outer eastern portion of the
image is presented in Fig.~\ref{fig:east}, to illustrate the quality of the
imaging away from bright sources.

As noted above, in the western part of the Lockman Hole field there is the
bright radio source 3C244.1 (e.g.\ \citealt{1987MNRAS.225....1A,
2004MNRAS.351..845G}), an extragalactic FR type II source
\citep{1974MNRAS.167P..31F} which is about 1~arcmin in extent, with an
integrated flux density of $\approx 10.5$~Jy at 610~MHz
(Fig.~\ref{fig:3C244p1}).

A catalogue of radio sources was made in a way similar to that used in Paper~I.
An initial catalogue of sources within the 30~per~cent primary response with a
peak brightness of greater than six times the local noise was made using {\sc
SExtractor} \citep{1996A&AS..117..393B}. Using the technique described in
\cite{2008MNRAS.383...75G}, close to brighter ($>10$~mJy peak) sources a more
stringent cutoff of at least 12 times the local noise was applied.
Table~\ref{tab:CLASXS} presents a sample of 163 sources from the catalogue, in
the region $10^{\rm h} 30^{\rm m} 50^{\rm s}$ to $10^{\rm h} 36^{\rm m} 20^{\rm
s}$, $+57^\circ 18'$ to $58^\circ 06'$, i.e.\ the region covered by the CLASXS
X-ray survey. Column 1 gives the IAU designation of the source, in the form
GMRTLH Jhhmmss.s+ddmmss, where J represents J2000.0 coordinates, hhmmss.s
represents RA in hours, minutes and truncated tenths of seconds and ddmmss
represents the Dec in degrees, arcminutes and truncated arcseconds. Columns 2
and 3 give the RA and Dec of the source, calculated using first moments of the
relevant pixel brightnesses to give a centroid position. Column 4 gives the
brightness of the peak pixel in each source, $S_{\rm peak}$, in
mJy~beam$^{-1}$, and Column 5 gives the local r.m.s.\ noise, $\sigma$, in
$\muup$Jy~beam$^{-1}$. Columns 6 and 7 give the integrated flux density and
error, $S_{\rm int}$ and $\Delta S_{\rm int}$, in mJy. Columns 8 and 9 give the
$X, Y$ pixel coordinates from the mosaic image of the source centroid. Column
10 is the {\sc SExtractor} deblended object flag: (1) where a nearby bright
source may be affecting the calculated flux, (2) where a source has been
deblended into two or more components from a single initial island of flux and
(3) where both of the above criteria apply. The final mosaic should be examined
to check the cases where one extended object has been represented by two or
more entries. Also, because of the limited dynamic range of the GMRT, the
parameters of sources near bright sources should be treated with some caution.

The final mosaicked image and the catalogue of 4934 sources are both available online
from {\tt http://www.mrao.cam.ac.uk/surveys/}, along with the results of our other GMRT
surveys of various SWIRE fields and the Spitzer Extragalactic First Look Survey region.

\section*{Acknowledgements}

We thank the staff of the GMRT who made these observations possible. In
particular we thank Nimisha Kantharia for making the re-scheduled observations.
The GMRT is run by the National Centre for Radio Astrophysics of the Tata
Institute of Fundamental Research.

%
\setlength{\bibsep}{0pt}
\def\newblock{}

\clearpage

\bgroup
%
%
\tablesize
\tabcolsep=3pt
\def\chead#1{\multicolumn{1}{c}{#1}}
\begin{center}
\begin{longtable}{lllrrrrrrr}
\caption{A sample of entries from the catalogue of radio sources in the Lockman
Hole region. See text for details. These are sources in the CLASXS
region.}\label{tab:CLASXS}\\
  \hline\hline
  \chead{Name} & \multicolumn{2}{c}{RA (J2000.0) Dec} & \chead{$S_{\rm peak}$}   & \chead{$\sigma$} & \chead{$S_{\rm int}$} & \chead{$\Delta S_{\rm int}$} & \chead{X}   & \chead{Y}   & \chead{flags} \\
  \chead{(1)}  & \chead{(2)}    & \chead{(3)}         & \chead{(4)}              & \chead{(5)}      & \chead{(6)}           & \chead{(7)}                  & \chead{(8)} & \chead{(9)} & \chead{(10)}  \\ \hline
\endfirsthead
  \hline
  \chead{Name} & \multicolumn{2}{c}{RA (J2000.0) Dec} & \chead{$S_{\rm peak}$}   & \chead{$\sigma$} & \chead{$S_{\rm int}$} & \chead{$\Delta S_{\rm int}$} & \chead{X} & \chead{Y} & \chead{flags} \\
  \chead{(1)}  & \chead{(2)}    & \chead{(3)}         & \chead{(4)}              & \chead{(5)}      & \chead{(6)}           & \chead{(7)}                  & \chead{(8)} & \chead{(9)} & \chead{(10)}  \\ \hline
\endhead
  \hline
\endfoot
  \hline\hline
\endlastfoot
 GMRTLH J103050.9+574657 & 10:30:50.99 & +57:46:57.0 &   0.874 &  140 &    1.043 &  0.126 & 11023 &  6097 & 0  \\
 GMRTLH J103055.6+572316 & 10:30:55.60 & +57:23:16.2 &   0.626 &   96 &    0.651 &  0.090 & 11048 &  5150 & 0  \\
 GMRTLH J103055.8+575948 & 10:30:55.82 & +57:59:48.9 &  14.804 &  154 &   19.687 &  0.330 & 10971 &  6610 & 0  \\
 GMRTLH J103059.3+572303 & 10:30:59.39 & +57:23:03.8 &   3.755 &  101 &    4.181 &  0.183 & 11028 &  5141 & 0  \\
 GMRTLH J103102.6+573805 & 10:31:02.65 & +57:38:05.6 &   1.591 &  121 &    1.466 &  0.143 & 10979 &  5740 & 0  \\
 GMRTLH J103105.0+575354 & 10:31:05.08 & +57:53:54.3 &  99.348 &  438 &  117.786 &  1.099 & 10934 &  6371 & 0  \\
 GMRTLH J103113.3+572242 & 10:31:13.30 & +57:22:42.2 &   0.552 &   87 &    0.969 &  0.100 & 10953 &  5122 & 0  \\
 GMRTLH J103118.5+574731 & 10:31:18.54 & +57:47:31.7 &   0.983 &  157 &    0.824 &  0.135 & 10875 &  6113 & 0  \\
 GMRTLH J103118.7+574758 & 10:31:18.78 & +57:47:58.0 &   1.129 &  160 &    0.972 &  0.185 & 10873 &  6130 & 0  \\
 GMRTLH J103120.3+571833 & 10:31:20.32 & +57:18:33.8 &   0.746 &  103 &    0.626 &  0.100 & 10924 &  4955 & 0  \\
 GMRTLH J103122.3+580211 & 10:31:22.31 & +58:02:11.7 &   1.087 &  150 &    2.705 &  0.229 & 10826 &  6697 & 0  \\
 GMRTLH J103123.0+574227 & 10:31:23.09 & +57:42:27.3 &   1.171 &  117 &    0.916 &  0.124 & 10861 &  5909 & 0  \\
 GMRTLH J103123.4+580559 & 10:31:23.46 & +58:05:59.3 &   1.201 &  192 &    2.040 &  0.198 & 10812 &  6849 & 0  \\
 GMRTLH J103126.1+573242 & 10:31:26.12 & +57:32:42.6 &   0.813 &  103 &    0.553 &  0.089 & 10864 &  5519 & 0  \\
 GMRTLH J103126.9+572328 & 10:31:26.94 & +57:23:28.7 &   0.504 &   80 &    0.392 &  0.069 & 10878 &  5149 & 0  \\
 GMRTLH J103129.5+572511 & 10:31:29.59 & +57:25:11.3 &   4.884 &   94 &    8.831 &  0.205 & 10861 &  5217 & 0  \\
 GMRTLH J103130.4+580340 & 10:31:30.43 & +58:03:40.3 &   1.109 &  165 &    3.122 &  0.262 & 10780 &  6754 & 0  \\
 GMRTLH J103131.2+573934 & 10:31:31.27 & +57:39:34.1 &   4.718 &  118 &    5.069 &  0.193 & 10823 &  5791 & 0  \\
 GMRTLH J103133.5+572040 & 10:31:33.53 & +57:20:40.6 &   0.702 &   90 &    0.518 &  0.081 & 10849 &  5036 & 0  \\
 GMRTLH J103134.3+574224 & 10:31:34.36 & +57:42:24.2 &   1.774 &  116 &    1.668 &  0.166 & 10801 &  5904 & 0  \\
 GMRTLH J103136.2+580024 & 10:31:36.27 & +58:00:24.1 &   1.240 &  163 &    2.901 &  0.238 & 10755 &  6622 & 0  \\
 GMRTLH J103138.3+575836 & 10:31:38.36 & +57:58:36.1 &   0.908 &  149 &    1.548 &  0.184 & 10748 &  6550 & 0  \\
 GMRTLH J103138.6+572625 & 10:31:38.64 & +57:26:25.6 &   5.531 &   92 &    5.861 &  0.157 & 10810 &  5264 & 0  \\
 GMRTLH J103141.8+580131 & 10:31:41.80 & +58:01:31.2 &   1.374 &  150 &    1.949 &  0.229 & 10724 &  6665 & 0  \\
 GMRTLH J103155.2+580518 & 10:31:55.25 & +58:05:18.4 &   1.463 &  205 &    2.919 &  0.265 & 10646 &  6813 & 0  \\
 GMRTLH J103156.2+580000 & 10:31:56.20 & +58:00:00.5 &   1.006 &  145 &    1.424 &  0.180 & 10651 &  6601 & 0  \\
 GMRTLH J103156.6+573845 & 10:31:56.62 & +57:38:45.6 &   1.741 &  104 &    1.704 &  0.140 & 10689 &  5752 & 0  \\
 GMRTLH J103157.0+580148 & 10:31:57.07 & +58:01:48.5 &   1.400 &  185 &    2.904 &  0.270 & 10643 &  6673 & 0  \\
 GMRTLH J103157.8+580137 & 10:31:57.80 & +58:01:37.5 &   1.155 &  190 &    1.775 &  0.203 & 10639 &  6665 & 0  \\
 GMRTLH J103200.3+580043 & 10:32:00.38 & +58:00:43.7 &   1.212 &  170 &    1.326 &  0.164 & 10627 &  6629 & 0  \\
 GMRTLH J103200.9+574944 & 10:32:00.95 & +57:49:44.3 &  19.057 &  181 &   20.071 &  0.370 & 10645 &  6190 & 0  \\
 GMRTLH J103201.4+580115 & 10:32:01.40 & +58:01:15.1 &   1.370 &  183 &    1.758 &  0.217 & 10621 &  6650 & 0  \\
 GMRTLH J103204.9+575309 & 10:32:04.94 & +57:53:09.1 &   0.982 &  149 &    1.444 &  0.180 & 10618 &  6325 & 0  \\
 GMRTLH J103207.6+572153 & 10:32:07.61 & +57:21:53.5 &  12.099 &   92 &   12.863 &  0.192 & 10663 &  5075 & 0  \\
 GMRTLH J103210.7+571821 & 10:32:10.71 & +57:18:21.5 &   1.583 &   82 &    1.442 &  0.108 & 10653 &  4933 & 0  \\
 GMRTLH J103215.6+580208 & 10:32:15.66 & +58:02:08.5 &   1.129 &  176 &    1.425 &  0.223 & 10544 &  6682 & 0  \\
 GMRTLH J103216.7+575435 & 10:32:16.72 & +57:54:35.4 &   1.052 &  157 &    1.087 &  0.151 & 10552 &  6380 & 0  \\
 GMRTLH J103220.1+575833 & 10:32:20.16 & +57:58:33.8 &   1.228 &  174 &    1.881 &  0.195 & 10527 &  6537 & 0  \\
 GMRTLH J103220.7+573835 & 10:32:20.72 & +57:38:35.0 &   3.483 &  102 &    3.837 &  0.163 & 10561 &  5739 & 0  \\
 GMRTLH J103221.7+573819 & 10:32:21.76 & +57:38:19.2 &   0.677 &  102 &    0.669 &  0.095 & 10556 &  5728 & 0  \\
 GMRTLH J103222.6+580234 & 10:32:22.63 & +58:02:34.4 &   1.499 &  242 &    3.274 &  0.265 & 10506 &  6697 & 0  \\
 GMRTLH J103222.8+575550 & 10:32:22.85 & +57:55:50.0 &   9.548 &  166 &   16.004 &  0.397 & 10518 &  6428 & 0  \\
 GMRTLH J103235.7+571859 & 10:32:35.72 & +57:18:59.6 &   0.547 &   83 &    0.464 &  0.071 & 10517 &  4953 & 0  \\
 GMRTLH J103238.3+572220 & 10:32:38.33 & +57:22:20.8 &   0.639 &   89 &    0.538 &  0.077 & 10496 &  5086 & 0  \\
 GMRTLH J103239.2+580046 & 10:32:39.29 & +58:00:46.2 &   1.317 &  216 &    2.256 &  0.262 & 10421 &  6621 & 0  \\
 GMRTLH J103239.8+574611 & 10:32:39.83 & +57:46:11.7 &   1.073 &  131 &    0.909 &  0.136 & 10445 &  6038 & 0  \\
 GMRTLH J103241.6+573010 & 10:32:41.60 & +57:30:10.5 &   2.310 &   93 &    2.541 &  0.143 & 10465 &  5398 & 0  \\
 GMRTLH J103241.6+580120 & 10:32:41.68 & +58:01:20.1 &   1.748 &  187 &    3.818 &  0.338 & 10408 &  6643 & 0  \\
 GMRTLH J103243.2+571809 & 10:32:43.29 & +57:18:09.0 &   0.509 &   79 &    0.366 &  0.064 & 10477 &  4917 & 0  \\
 GMRTLH J103243.2+580029 & 10:32:43.24 & +58:00:29.1 &   1.395 &  212 &    1.899 &  0.290 & 10401 &  6609 & 0  \\
 GMRTLH J103252.5+573115 & 10:32:52.55 & +57:31:15.5 &   7.251 &  108 &   11.178 &  0.242 & 10404 &  5438 & 3  \\
 GMRTLH J103254.6+573124 & 10:32:54.68 & +57:31:24.7 &  11.001 &  102 &   14.394 &  0.230 & 10392 &  5444 & 3  \\
 GMRTLH J103259.8+575321 & 10:32:59.86 & +57:53:21.1 &   1.501 &  126 &    1.276 &  0.142 & 10326 &  6320 & 0  \\
 GMRTLH J103303.3+575517 & 10:33:03.39 & +57:55:17.7 &   1.096 &  139 &    6.331 &  0.320 & 10304 &  6396 & 0  \\
 GMRTLH J103305.9+573502 & 10:33:05.92 & +57:35:02.5 &   0.702 &  103 &    0.503 &  0.084 & 10326 &  5587 & 0  \\
 GMRTLH J103311.1+574208 & 10:33:11.16 & +57:42:08.4 &   1.035 &  103 &    1.898 &  0.153 & 10285 &  5869 & 3  \\
 GMRTLH J103312.3+574217 & 10:33:12.35 & +57:42:17.1 &   1.278 &  102 &    2.416 &  0.178 & 10279 &  5874 & 2  \\
 GMRTLH J103314.5+573935 & 10:33:14.51 & +57:39:35.0 &   3.740 &  104 &    5.446 &  0.205 & 10272 &  5766 & 3  \\
 GMRTLH J103314.6+573920 & 10:33:14.61 & +57:39:20.6 &   6.056 &  103 &    7.879 &  0.204 & 10272 &  5756 & 3  \\
 GMRTLH J103315.5+573101 & 10:33:15.56 & +57:31:01.9 &   0.763 &   91 &    0.720 &  0.097 & 10281 &  5424 & 0  \\
 GMRTLH J103317.3+572718 & 10:33:17.39 & +57:27:18.6 &   0.581 &   82 &    0.386 &  0.070 & 10278 &  5275 & 0  \\
 GMRTLH J103324.8+575931 & 10:33:24.85 & +57:59:31.5 &   1.116 &  177 &    1.692 &  0.228 & 10182 &  6561 & 0  \\
 GMRTLH J103328.1+573241 & 10:33:28.19 & +57:32:41.3 &   0.619 &   85 &    0.586 &  0.080 & 10210 &  5487 & 0  \\
 GMRTLH J103332.7+580012 & 10:33:32.75 & +58:00:12.2 &   1.218 &  179 &    1.912 &  0.212 & 10139 &  6586 & 1  \\
 GMRTLH J103332.9+580028 & 10:33:32.94 & +58:00:28.2 &   1.104 &  181 &    3.657 &  0.264 & 10138 &  6596 & 0  \\
 GMRTLH J103333.8+575947 & 10:33:33.82 & +57:59:47.3 &   1.427 &  184 &    1.937 &  0.222 & 10135 &  6569 & 0  \\
 GMRTLH J103334.1+573201 & 10:33:34.18 & +57:32:01.0 &   0.504 &   81 &    0.680 &  0.076 & 10179 &  5459 & 0  \\
 GMRTLH J103343.2+572549 & 10:33:43.29 & +57:25:49.7 &   0.530 &   82 &    0.413 &  0.067 & 10141 &  5210 & 0  \\
 GMRTLH J103344.6+573332 & 10:33:44.60 & +57:33:32.6 &   0.528 &   79 &    0.337 &  0.061 & 10121 &  5518 & 0  \\
 GMRTLH J103345.9+574013 & 10:33:45.99 & +57:40:13.0 &   0.540 &   84 &    0.524 &  0.069 & 10103 &  5784 & 0  \\
 GMRTLH J103346.6+575504 & 10:33:46.64 & +57:55:04.1 &   0.766 &  126 &    0.758 &  0.098 & 10074 &  6378 & 0  \\
 GMRTLH J103347.0+574945 & 10:33:47.02 & +57:49:45.9 &   0.721 &  103 &    0.635 &  0.084 & 10081 &  6166 & 0  \\
 GMRTLH J103357.7+573653 & 10:33:57.73 & +57:36:53.9 &   1.108 &   95 &    0.787 &  0.099 & 10045 &  5649 & 0  \\
 GMRTLH J103358.7+574316 & 10:33:58.76 & +57:43:16.2 &   2.082 &   90 &    2.152 &  0.134 & 10029 &  5903 & 0  \\
 GMRTLH J103359.2+572951 & 10:33:59.20 & +57:29:51.6 &   2.956 &   76 &    3.249 &  0.137 & 10049 &  5367 & 0  \\
 GMRTLH J103405.4+573329 & 10:34:05.42 & +57:33:29.1 &   0.477 &   70 &    0.910 &  0.076 & 10010 &  5511 & 0  \\
 GMRTLH J103408.5+573702 & 10:34:08.52 & +57:37:02.3 &   0.598 &   83 &    0.513 &  0.075 &  9987 &  5652 & 0  \\
 GMRTLH J103411.5+575528 & 10:34:11.50 & +57:55:28.5 &   0.887 &  145 &    0.917 &  0.145 &  9942 &  6388 & 0  \\
 GMRTLH J103414.9+573219 & 10:34:14.91 & +57:32:19.9 &   0.530 &   82 &    0.736 &  0.082 &  9961 &  5463 & 0  \\
 GMRTLH J103415.4+573408 & 10:34:15.41 & +57:34:08.1 &   1.546 &   75 &    1.695 &  0.126 &  9955 &  5535 & 0  \\
 GMRTLH J103419.7+574203 & 10:34:19.72 & +57:42:03.5 &   0.585 &   91 &    1.057 &  0.115 &  9919 &  5850 & 0  \\
 GMRTLH J103422.4+574207 & 10:34:22.46 & +57:42:07.9 &   1.277 &  103 &    3.926 &  0.226 &  9905 &  5853 & 0  \\
 GMRTLH J103423.5+572136 & 10:34:23.54 & +57:21:36.0 &   1.099 &   79 &    1.073 &  0.096 &  9931 &  5032 & 0  \\
 GMRTLH J103425.5+574212 & 10:34:25.55 & +57:42:12.8 &   0.855 &  102 &    1.289 &  0.126 &  9888 &  5855 & 0  \\
 GMRTLH J103426.7+574254 & 10:34:26.76 & +57:42:54.6 &   0.653 &   90 &    0.597 &  0.080 &  9881 &  5883 & 0  \\
 GMRTLH J103433.3+573146 & 10:34:33.33 & +57:31:46.6 &   0.847 &   70 &    0.894 &  0.085 &  9863 &  5437 & 0  \\
 GMRTLH J103434.1+572910 & 10:34:34.17 & +57:29:10.5 &   0.814 &   76 &    1.005 &  0.107 &  9862 &  5332 & 0  \\
 GMRTLH J103435.6+572759 & 10:34:35.68 & +57:27:59.6 &   1.395 &   84 &    1.520 &  0.117 &  9856 &  5285 & 0  \\
 GMRTLH J103436.3+575048 & 10:34:36.32 & +57:50:48.7 &   0.817 &  105 &    0.595 &  0.090 &  9817 &  6197 & 0  \\
 GMRTLH J103446.9+573105 & 10:34:46.97 & +57:31:05.6 &   0.568 &   91 &    0.321 &  0.062 &  9791 &  5407 & 0  \\
 GMRTLH J103450.3+573949 & 10:34:50.33 & +57:39:49.4 &   0.994 &   84 &    0.748 &  0.090 &  9759 &  5755 & 0  \\
 GMRTLH J103452.7+580441 & 10:34:52.74 & +58:04:41.0 &   1.702 &  261 &    2.359 &  0.243 &  9709 &  6748 & 0  \\
 GMRTLH J103452.9+580426 & 10:34:52.90 & +58:04:26.6 &   1.630 &  253 &    5.148 &  0.326 &  9709 &  6738 & 0  \\
 GMRTLH J103453.7+575405 & 10:34:53.79 & +57:54:05.2 &   0.730 &  117 &    0.615 &  0.101 &  9720 &  6324 & 0  \\
 GMRTLH J103454.1+580507 & 10:34:54.19 & +58:05:07.9 &   2.380 &  327 &    3.529 &  0.413 &  9701 &  6766 & 0  \\
 GMRTLH J103454.7+573318 & 10:34:54.75 & +57:33:18.9 &   0.979 &  120 &    1.283 &  0.139 &  9746 &  5494 & 0  \\
 GMRTLH J103455.2+574518 & 10:34:55.28 & +57:45:18.1 &   0.561 &   90 &    0.488 &  0.087 &  9725 &  5973 & 0  \\
 GMRTLH J103457.1+580517 & 10:34:57.10 & +58:05:17.2 &   3.392 &  326 &    8.473 &  0.588 &  9685 &  6771 & 0  \\
 GMRTLH J103457.2+580253 & 10:34:57.21 & +58:02:53.1 &   1.464 &  190 &    3.343 &  0.306 &  9688 &  6675 & 0  \\
 GMRTLH J103457.9+573941 & 10:34:57.92 & +57:39:41.3 &   0.626 &   85 &    0.707 &  0.082 &  9719 &  5748 & 0  \\
 GMRTLH J103459.4+573256 & 10:34:59.49 & +57:32:56.9 &   0.807 &  115 &    0.508 &  0.089 &  9721 &  5478 & 0  \\
 GMRTLH J103500.2+580458 & 10:35:00.26 & +58:04:58.7 &   1.850 &  241 &    2.477 &  0.278 &  9669 &  6758 & 0  \\
 GMRTLH J103501.2+573317 & 10:35:01.28 & +57:33:17.8 &   0.811 &  128 &    0.823 &  0.105 &  9711 &  5492 & 0  \\
 GMRTLH J103501.7+571930 & 10:35:01.70 & +57:19:30.8 &   0.495 &   78 &    0.378 &  0.064 &  9728 &  4941 & 0  \\
 GMRTLH J103502.3+572341 & 10:35:02.36 & +57:23:41.2 &  11.880 &   98 &   12.775 &  0.204 &  9719 &  5107 & 0  \\
 GMRTLH J103504.0+573128 & 10:35:04.02 & +57:31:28.0 &   0.858 &  137 &    0.790 &  0.112 &  9699 &  5418 & 0  \\
 GMRTLH J103505.1+575821 & 10:35:05.15 & +57:58:21.5 &   1.039 &  166 &    1.020 &  0.142 &  9653 &  6493 & 0  \\
 GMRTLH J103505.7+572737 & 10:35:05.79 & +57:27:37.8 &   1.712 &   78 &    1.991 &  0.125 &  9695 &  5264 & 0  \\
 GMRTLH J103505.9+573030 & 10:35:05.92 & +57:30:30.1 &   0.694 &  111 &    0.537 &  0.091 &  9690 &  5379 & 0  \\
 GMRTLH J103506.0+580422 & 10:35:06.02 & +58:04:22.8 &   1.639 &  250 &    2.120 &  0.274 &  9640 &  6733 & 0  \\
 GMRTLH J103507.9+574227 & 10:35:07.94 & +57:42:27.6 &   0.524 &   81 &    0.468 &  0.067 &  9662 &  5857 & 0  \\
 GMRTLH J103508.0+580409 & 10:35:08.04 & +58:04:09.7 &   2.457 &  265 &    4.767 &  0.415 &  9629 &  6724 & 0  \\
 GMRTLH J103508.8+573737 & 10:35:08.88 & +57:37:37.7 &   1.449 &   92 &    1.175 &  0.109 &  9664 &  5663 & 0  \\
 GMRTLH J103509.2+573302 & 10:35:09.29 & +57:33:02.9 &   0.949 &  155 &    0.684 &  0.120 &  9668 &  5480 & 0  \\
 GMRTLH J103509.7+580230 & 10:35:09.77 & +58:02:30.4 &   1.486 &  222 &    1.532 &  0.214 &  9623 &  6658 & 2  \\
 GMRTLH J103510.2+580444 & 10:35:10.26 & +58:04:44.8 &   2.572 &  263 &    3.638 &  0.352 &  9617 &  6747 & 0  \\
 GMRTLH J103511.3+573403 & 10:35:11.38 & +57:34:03.4 &   0.906 &  139 &    1.260 &  0.175 &  9655 &  5520 & 0  \\
 GMRTLH J103511.5+580307 & 10:35:11.55 & +58:03:07.4 &   1.700 &  213 &    1.996 &  0.246 &  9612 &  6682 & 0  \\
 GMRTLH J103511.8+573222 & 10:35:11.85 & +57:32:22.9 &   0.926 &  150 &    0.871 &  0.129 &  9655 &  5453 & 0  \\
 GMRTLH J103512.2+573125 & 10:35:12.25 & +57:31:25.5 &   1.018 &  148 &    0.732 &  0.121 &  9654 &  5415 & 0  \\
 GMRTLH J103512.8+574001 & 10:35:12.83 & +57:40:01.4 &   0.602 &   80 &    0.581 &  0.077 &  9639 &  5758 & 0  \\
 GMRTLH J103513.4+580547 & 10:35:13.46 & +58:05:47.9 &   3.714 &  455 &   19.052 &  1.124 &  9598 &  6789 & 0  \\
 GMRTLH J103513.7+580454 & 10:35:13.79 & +58:04:54.0 &   1.851 &  278 &    1.657 &  0.238 &  9598 &  6753 & 0  \\
 GMRTLH J103513.9+574249 & 10:35:13.93 & +57:42:49.8 &   0.706 &   83 &    0.533 &  0.080 &  9629 &  5870 & 0  \\
 GMRTLH J103514.6+573142 & 10:35:14.62 & +57:31:42.3 &   1.044 &  151 &    0.955 &  0.123 &  9641 &  5426 & 0  \\
 GMRTLH J103516.2+580146 & 10:35:16.25 & +58:01:46.9 &   1.744 &  213 &    3.388 &  0.330 &  9589 &  6627 & 0  \\
 GMRTLH J103516.5+580057 & 10:35:16.51 & +58:00:57.5 &   0.980 &  157 &    2.729 &  0.190 &  9589 &  6594 & 0  \\
 GMRTLH J103517.9+580517 & 10:35:17.91 & +58:05:17.8 &   2.337 &  346 &    6.311 &  0.564 &  9576 &  6768 & 0  \\
 GMRTLH J103520.6+580123 & 10:35:20.61 & +58:01:23.1 &   1.335 &  197 &    1.829 &  0.209 &  9567 &  6611 & 3  \\
 GMRTLH J103521.5+574840 & 10:35:21.50 & +57:48:40.3 &   0.529 &   86 &    0.392 &  0.071 &  9580 &  6102 & 0  \\
 GMRTLH J103522.1+573720 & 10:35:22.14 & +57:37:20.0 &   1.997 &   96 &    2.099 &  0.136 &  9593 &  5649 & 0  \\
 GMRTLH J103523.2+573900 & 10:35:23.24 & +57:39:00.1 &   4.368 &   85 &    4.407 &  0.143 &  9585 &  5715 & 0  \\
 GMRTLH J103523.3+573247 & 10:35:23.32 & +57:32:47.7 &   3.619 &  128 &    8.312 &  0.307 &  9593 &  5467 & 0  \\
 GMRTLH J103524.0+573526 & 10:35:24.06 & +57:35:26.7 &   0.655 &   95 &    0.513 &  0.081 &  9585 &  5573 & 0  \\
 GMRTLH J103526.9+573035 & 10:35:26.90 & +57:30:35.2 &   2.264 &  129 &    2.555 &  0.183 &  9577 &  5378 & 0  \\
 GMRTLH J103528.3+580320 & 10:35:28.37 & +58:03:20.3 &   1.534 &  249 &    3.334 &  0.301 &  9523 &  6687 & 0  \\
 GMRTLH J103529.2+572925 & 10:35:29.28 & +57:29:25.5 &   0.694 &  101 &    0.487 &  0.078 &  9566 &  5332 & 0  \\
 GMRTLH J103530.3+575216 & 10:35:30.38 & +57:52:16.9 &   0.867 &  115 &    1.175 &  0.130 &  9528 &  6245 & 0  \\
 GMRTLH J103533.0+573259 & 10:35:33.08 & +57:32:59.8 &   0.821 &  118 &    0.520 &  0.091 &  9540 &  5474 & 0  \\
 GMRTLH J103534.6+572814 & 10:35:34.62 & +57:28:14.6 &   0.631 &   99 &    0.579 &  0.089 &  9539 &  5283 & 0  \\
 GMRTLH J103536.8+573329 & 10:35:36.89 & +57:33:29.2 &   0.694 &  110 &    1.010 &  0.114 &  9519 &  5492 & 0  \\
 GMRTLH J103540.1+573324 & 10:35:40.19 & +57:33:24.5 &   0.871 &  102 &    1.413 &  0.150 &  9502 &  5489 & 0  \\
 GMRTLH J103540.5+575529 & 10:35:40.56 & +57:55:29.7 &   0.746 &  123 &    1.067 &  0.114 &  9469 &  6372 & 0  \\
 GMRTLH J103541.3+575622 & 10:35:41.35 & +57:56:22.1 &   0.816 &  132 &    1.385 &  0.132 &  9464 &  6406 & 0  \\
 GMRTLH J103541.9+580415 & 10:35:41.94 & +58:04:15.6 &   1.026 &  162 &    1.833 &  0.217 &  9450 &  6722 & 0  \\
 GMRTLH J103542.3+580441 & 10:35:42.30 & +58:04:41.4 &   1.077 &  163 &    1.004 &  0.140 &  9448 &  6739 & 0  \\
 GMRTLH J103544.8+573206 & 10:35:44.81 & +57:32:06.9 &   0.681 &   88 &    0.714 &  0.097 &  9479 &  5436 & 3  \\
 GMRTLH J103546.1+574310 & 10:35:46.16 & +57:43:10.4 &   0.818 &   82 &    0.938 &  0.093 &  9457 &  5878 & 0  \\
 GMRTLH J103550.4+573629 & 10:35:50.48 & +57:36:29.8 &   0.648 &   91 &    0.578 &  0.078 &  9442 &  5610 & 0  \\
 GMRTLH J103550.6+573257 & 10:35:50.67 & +57:32:57.5 &   0.569 &   85 &    0.624 &  0.085 &  9446 &  5469 & 0  \\
 GMRTLH J103554.3+574642 & 10:35:54.30 & +57:46:42.0 &   0.556 &   88 &    0.488 &  0.072 &  9408 &  6018 & 0  \\
 GMRTLH J103557.4+573517 & 10:35:57.43 & +57:35:17.3 &   0.770 &   77 &    0.611 &  0.099 &  9407 &  5561 & 0  \\
 GMRTLH J103559.1+572429 & 10:35:59.16 & +57:24:29.9 &   0.515 &   85 &    0.650 &  0.090 &  9412 &  5129 & 0  \\
 GMRTLH J103600.8+580048 & 10:36:00.88 & +58:00:48.3 &   1.132 &  185 &    3.704 &  0.294 &  9355 &  6580 & 0  \\
 GMRTLH J103607.3+575106 & 10:36:07.33 & +57:51:06.0 &   3.120 &  132 &    3.661 &  0.221 &  9333 &  6191 & 0  \\
 GMRTLH J103609.2+573245 & 10:36:09.25 & +57:32:45.6 &   0.624 &   85 &    0.941 &  0.094 &  9347 &  5458 & 0  \\
 GMRTLH J103610.0+575115 & 10:36:10.00 & +57:51:15.8 &  18.114 &  137 &   25.505 &  0.343 &  9319 &  6197 & 0  \\
 GMRTLH J103610.4+573134 & 10:36:10.42 & +57:31:34.7 &   0.516 &   83 &    0.479 &  0.077 &  9342 &  5410 & 0  \\
 GMRTLH J103611.8+572524 & 10:36:11.87 & +57:25:24.3 &   0.529 &   83 &    0.738 &  0.080 &  9342 &  5163 & 0  \\
 GMRTLH J103613.6+574004 & 10:36:13.69 & +57:40:04.7 &   0.553 &   83 &    0.597 &  0.081 &  9314 &  5749 & 0  \\
 GMRTLH J103614.4+572346 & 10:36:14.46 & +57:23:46.9 &   0.593 &   85 &    0.703 &  0.076 &  9331 &  5098 & 0  \\
 GMRTLH J103614.4+573554 & 10:36:14.41 & +57:35:54.3 &   0.633 &   95 &    0.417 &  0.074 &  9315 &  5582 & 0  \\
 GMRTLH J103617.6+572617 & 10:36:17.68 & +57:26:17.3 &   0.630 &   87 &    0.981 &  0.096 &  9310 &  5197 & 0  \\
\end{longtable}
\end{center}
\egroup

\label{lastpage}
\end{document}